\documentclass[9pt,twocolumn,twoside,lineno]{pnas-new}

\templatetype{pnasresearcharticle} 

\title{Listen to what they say: Better understand and detect online misinformation with user feedback} 

\author[a,b]{Hubert Etienne}
\author[a]{Onur Çelebi} 

\affil[a]{Facebook AI Research}
\affil[b]{Ecole Normale Supérieure}

\leadauthor{Etienne}

\keywords{Misinformation $|$ False news $|$ Reporting $|$ Facebook $|$ Instagram $|$} 

\begin{abstract}
  Social media users who report content are key allies in the management of online misinformation; however, no research has been conducted yet to understand their role and the different trends underlying their reporting activity. We suggest an original approach to studying misinformation: examining it from the reporting users’ perspective at the content-level and comparatively across regions and platforms. We propose the first classification of reported content pieces, resulting from a review of c. 9,000 items reported on Facebook and Instagram in France, the UK, and the US in June 2020. This allows us to observe meaningful distinctions regarding reporting content between countries and platforms as it significantly varies in volume, type, topic, and manipulation technique. Examining six of these techniques, we identify a novel one that is specific to Instagram US and significantly more sophisticated than others, potentially presenting a concrete challenge for algorithmic detection and human moderation. We also identify four reporting behaviours, from which we derive four types of noise capable of explaining half of the inaccuracy found in content reported as misinformation. We finally show that breaking down the user reporting signal into a plurality of behaviours allows to train a simple, although competitive, classifier on a small dataset with a combination of basic users-reports to classify the different types of reported content pieces.\end{abstract}

\begin{document}

\maketitle
\thispagestyle{firststyle}
\ifthenelse{\boolean{shortarticle}}{\ifthenelse{\boolean{singlecolumn}}{\abscontentformatted}{\abscontent}}{}

\dropcap{S}ocial media users who report content (thereafter abbreviated ‘reporters’) are key allies in the management of online misinformation. By posting comments expressing disbelief and providing fact-checking materials, they constitute the first line of defence against the potential virality of a hoax on social media and reduce the impact of false news on people’s beliefs. By reporting on these claims and encouraging others to do so, reporters also provide moderators with relevant signals supporting misinformation detection. However, social media users’ reports are often considered noisy signals, complex to integrate into algorithmic detection models supporting the prioritisation of relevant content for fact-checking. 
\footnote{Detection models leverage many signals to attribute a prevalence score to a given content item. User reports are only one of these and this paper’s scope is limited to better understand it.} This explains why reporters are still absent from the growing literature on misinformation, which instead focuses on those who spread hoaxes. Another gap in the literature relates to the lack of web-data-based comparative analysis between countries and platforms, although such research is necessary to develop an in-depth understanding of misinformation. We suggest an original approach to fill some of these gaps: leveraging mixed methods to examine misinformation from the reporters’ perspective, at the content-level (content marked by reporters as misinformation) comparatively across regions and platforms.\medskip

This paper aims to demonstrate the relevance of such an approach for improving the understanding of misinformation on social media and developing better moderation methods. Its contribution to this objective is threefold. First, it proposes a general classification of reported content (GCRC) resulting from the human review of c. 9,000 content items reported on Facebook and Instagram in France, the UK, and the US in June 2020. The methodology provided is extensively detailed to enable future studies to use it. This allows us to draw meaningful distinctions between countries and temper the discourse on a global ‘infodemic’
\footnote{https://www.who.int/news/item/23-09-2020-managing-the-covid-19-infodemic-promoting-healthy-behaviours-and-mitigating-the-harm-from-misinformation-and-disinformation.}; for example, it seems that misinformation on Covid-19 did not strike France as severely it did the US, differing in both volume, type, topics, and manipulation techniques. Second, in addition to five traditional information manipulation techniques, the study identifies a novel one, which is specific to Instagram US and presents significantly more challenges for algorithmic detection and human moderation. Third, it suggests four reporting behaviours that can explain the majority of the inaccuracy (55\%) in reporting. Breaking down the inaccuracy into four types of noise associated to these different behaviours, we show how a gradient boosting classification model trained on a combination of user reports can accurately classify these types of noise.

\section*{A reporter-oriented approach to studying misinformation at the content-level}

\subsection*{General approach}

Misinformation research benefits from a diversity of methodologies, selected countries, and social platforms. This, however, makes it difficult to compare results and generalise findings. Furthermore, with a few exceptions \cite{humprecht2020resilience,cinelli2021echo}), the selection of several countries and platforms in web-data-based research is justified by data augmentation purposes, rather than for comparative analysis. The resulting taxonomies of misinformation content \cite{wardle20166,molina2021fake,innes2020techniques} thus do not provide information on specific platforms and countries, whereas the annual Reuters Institute Digital News Report suggests that misinformation is significantly sensible to these variables. Our comparative approach aims to contribute to this effort, differentiating misinformation manifestations and practices across platforms and regions for which existing research is abundant (Facebook, especially the US and the UK) and those for which it is minimal or inexistant (Instagram, especially France). While the proportion of people using social networks to access the news has been relatively stable across Facebook, Twitter, and YouTube (+0\% +4.9\% +4.6\% CAGR), this figure has increased five-and-a-half-fold for Instagram between 2014-2020 \cite{newman2020reuters}. We thus expect this latter platform to provide a better observation point to arrive to new insights on misinformation types and practices.

The high volume of misinformation content and the limited access to relevant datasets lead most researchers to study misinformation indirectly such as via surveys or at the aggregated-data level. Surveys are valuable to capture people’s general sentiments about misinformation \cite{newman2020reuters} and how perceived misinformation impacts their trust in information sources \cite{altay2019so}. However, they are limited by the respondents’ memory, sincerity, and ability to detect hoaxes \cite{barthel2016many}. Aggregated data analyses are valuable to verify a hypothesis using large datasets. However, they do not allow content-level observations and are subject to major methodological limitations. These include a strong dependency on fact-checkers’ ratings, who have their own guidelines (as not all false news call for moderation), only review items reaching specific virality thresholds (in terms of engagement and impressions) and whose ratings serve as feedback to train detection algorithms (responsible for enqueuing relevant content for fact-checking). We aim to overcome these limitations by observing misinformation at the closest level, conducting a human review of reported content without any pre-selection based on relevance criteria, and establishing our own classification to analyse it. We thereby built a dataset reflecting the users’ perception of misinformation.

\begin{figure}[h]
\centering
\includegraphics[width=1.0\linewidth]{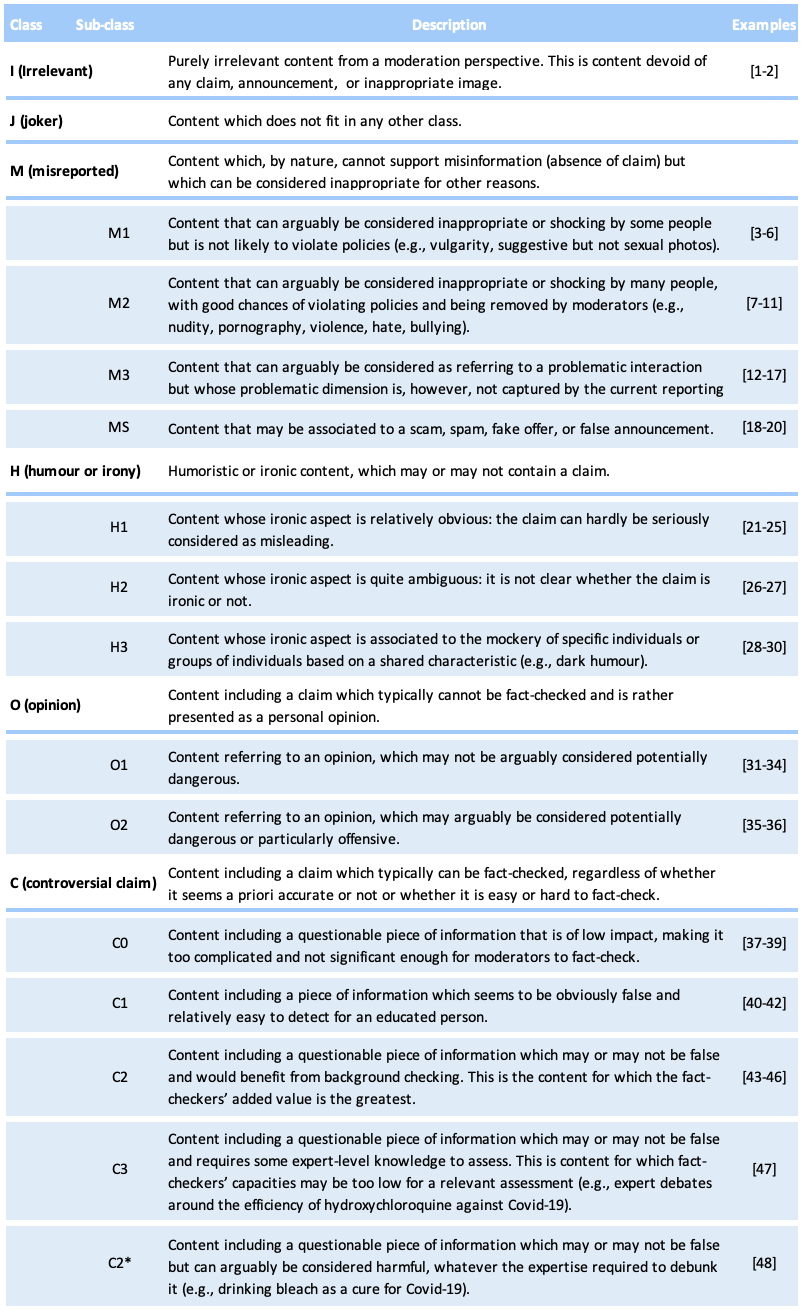}
\caption{The General classification of reported content (GCRC).}
\label{fig:fig1}
\end{figure}

\subsection*{A general classification for reported content}

The review and labelling of a sample S1 of c. 9,000 content items reported on Facebook and Instagram in France, the UK, and the US in June 2020 allowed us to propose the general classification of reported content (GCRC) below. Information related to S1, the review methodology and how it allowed us to address the four main difficulties we identified in labelling reported content are presented in Annex 1.

\vspace{-0.1cm}	

\section*{Country and platform specificities of content reported as misinformation}

Previous research has observed that misinformation topics may vary with countries \cite{humprecht2019fake} and that people express concern about different types of content across regions \cite{newman2020reuters}. We find that false news may also greatly differs in volume, type, and technique used between countries and platforms.

\subsection*{Reported content varies by volume and type across regions and platforms}

Figures 2 illustrate the distributions of S1 per class and country. The green line represents the normalised distribution of reported content per class. The blue bars represent the deviation of number of reports between the two platforms for the same country.
\vspace{-0.2cm}	
\begin{figure}[H]
\centering
\includegraphics[width=1\linewidth]{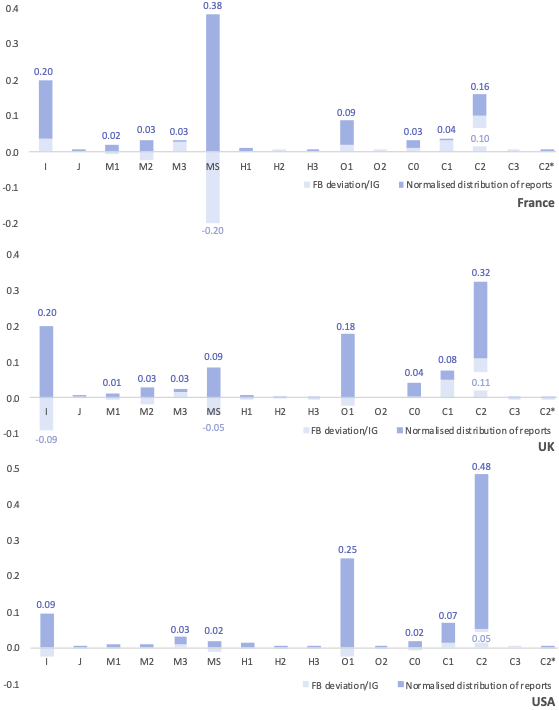}
\caption{Distribution of reports per class and platforms for each country}
\label{fig:fig2}
\end{figure}
\vspace{-0.2cm}	

A first observation is that the relative homogeneity in M1, M2, and M3 content across countries breaks at the platform-level in France and the UK. M3 content seems more specific to Facebook with 159 items in FB FR \& FB UK versus 36 for IG FR \& IG UK, while M1 and M2 are more specific to Instagram (resp. 214 in IG FR \& IG UK vs. 60 in FB FR \& FB UK). This may be explained by the fact that M3 mostly refer to denunciations, warnings, and calls for support, which better fit semantic posts than pictural ones. In contrast, most M1 and M2 items represents violent, offending, or sexually suggestive content, for which pictural posts are better suited.\medskip

A second point relates to the great difference in the volume of controversial content reported. C1 and C2 are 1.8x less numerous in FB FR than in FB UK and 2.0x less numerous compared to FB US. This trend is even more apparent on Instagram (resp. 3.6x and 7.3x lower) where only 4 items were classified as C1 in France (vs. 40 on IG UK and 86 on IG US). This could be explained either by a lower (a) volume of false news circulating in France, (b) reporting activity from French users, or (c) capacity or willingness of French users to report C content. Hypothesis (b) should be rejected, at least for Instagram, where the number of distinct content items reported per a thousand monthly active users is similar across the countries (0.78 FR, 0.81 UK, and 0.81 US). Hypothesis (c) is also hard to defend considering that 950 of the 1,492 reported content items in IG FR were associated to some kind of policy offense (vs. 186 for IG UK and 133 for IG US), suggesting great seriousness in French reporting. Humprecht et al. \cite{humprecht2020resilience} also recently suggested that French people may be four times more resilient to online misinformation than Americans. Finally, the qualitative analysis of S1 provides another argument supporting hypothesis (a) over (c). Whereas the great majority of C2 content in the US subsets was very likely to be a serious hoax, C2 items in FR subsets were much less likely to express false news. While they were mostly classified as C2 because of our reporter’s best intention (RBI) assumption (see Annex 1), they seemed to come from a general scepticism towards the government, especially concerning the management of masks, which was at the centre of a political scandal in June 2020 \cite{Moullot2020}.\medskip

Despite similar reporting rates and polarised contexts (Covid-19 restrictions and anti-police riots happening in the three countries), reported content was significantly smaller in both volume and severity in France than in the US, with the UK occupying an in-between position. While they were seem to have not been really concerned by misinformation issues, French Instagram users experienced a different issue due to the significantly reported use of spam, with MS items accounting for 58\% of IG FR. Lastly, in France, and to a lesser extent the UK, there are great differences in the distribution of reported content per classes between the two platforms. It is clear that C content is predominant on Facebook, rather than Instagram. Such variations, however, tend to disappear in the US, where the great symmetry in reported content types suggests an increasing uniformization of content on both platforms, therefore that such a belief is not accurate anymore.

\subsection*{Topics of reported content vary by region and platform}

\begin{figure}[h]
\centering
\includegraphics[width=1\linewidth]{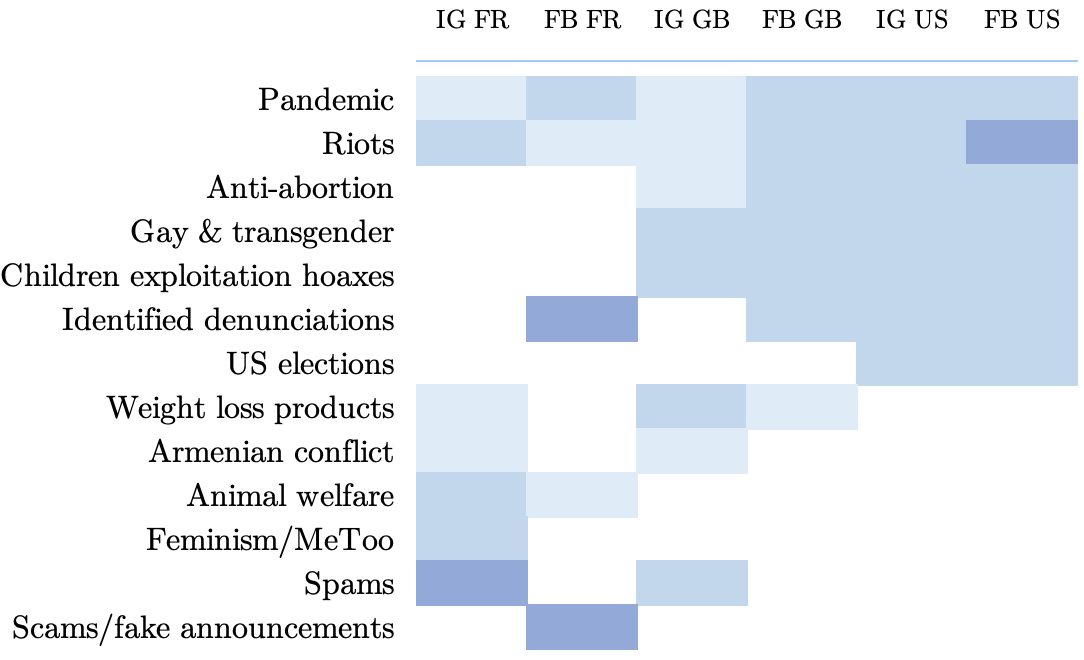}
\caption{Qualitative representation of S1 subsets’ topics}
\label{fig:fig3}
\end{figure}

Figure \ref{fig:fig3} shows the main topics expressed in reported content across S1 subsets. The qualitative scale ranges from light blue (a few items), to medium blue (a significant number of items), to dark blue (a great number of items). \medskip

Our review confirms that there are significant variations across regions in the topics considered polemical (as reported) by users. Less intuitively, we observe that the relative homogeneity in content between the two platforms in the US decreases in the UK and disappears in France. Besides the two universal topics present in all subsets (the pandemic and riots), only one other topic is present in both FB FR and IG FR, namely animal welfare, and it is not even expressed in the same way–IG posts express political claims related to animal rights, whereas FB posts only denounce particular cases of animal cruelty. In contrast, there are similar concerns about weight loss products and the Armenian conflict in IG FR and IG UK. Furthermore, even when users’ activity is monopolised by common topics associated to global events, it does not necessarily focus on the same aspects–e.g., while most controversial posts in FB US and IG US relate to serious hoaxes about masks, vaccines, and the reality of the pandemic, those from FB FR and IG FR generally criticise the government’s restrictions and not science.\medskip

Another finding is that most of the C2 items from FR and UK that were confirmed to contain hoaxes by fact-checkers (what we label VM) were concentrated into a narrow cluster of identical posts (n < 10) replicated dozens of times. This contrasts with the US subsets, where C2 posts with a high chance of being hoaxes are highly diversified. We also found a number of identical controversial posts in both the FB US and FB UK samples and to a lesser extent in the FB FR. These are likely coming from the US, often related to Covid-19 but also to US-centred events (BLM riots, including local events such as hoaxes about vandalised statues and cemeteries). In contrast, no posts related to French or English topics were found in the US samples. This suggests that several hoaxes are transmitted from the US to Europe and better circulate between countries on the same platform rather than between platforms in the same country. This was not obvious considering that language is expected to be a strong barrier to content circulation, all reporters of S1 content also have a Facebook account linked to their Instagram account, and many IG US C2 items are suspicious screenshots of Twitter posts.

\subsection*{Manipulation techniques vary by region and platform: the rising complexity of Instagram}

We identified six main techniques supporting misinformation associated to four psychological leverages: scepticism ((1)\textit{the revelation} and (2) \textit{the critical tipping point}), pragmatism ((3) \textit{false facts supported by false evidence}, (4) \textit{misleading presentation of facts}), empathy ((5) \textit{the confusion of feelings}) and coolness ((6) \textit{the excuse of casualness}). They are presented in Annex 2. While the first five techniques are consistent with existing typologies \cite{wardle20166,Yurkova2018,molina2021fake,innes2020techniques}, \textit{the excuse of casualness }seems both recent and specific to Instagram, especially IG US, emerging from the type of content shared on the platform. It characterises items with humour (jokes, ironical statements, memes, caricatures) and/or an artistic dimension (cartoons, short format videos, songs), making more or less explicit reference and reacting to a sub-claim using a frivolous tone. This content differs from both ‘parodies’ \cite{wardle20166} and ‘satires’ as it is not ‘meant to be perceived as unrealistic’ \cite{molina2021fake} but rather hides an assumed claim behind a veil of frivolity. Based on suggestion and sarcastic humour, this content does not introduce a new hoax but rather may constitute an efficient second-layer relay to support a hoax’s viral propagation. The apparent lack of seriousness makes the content both harder to detect and to moderate. \medskip
\vspace{-0.03cm}	

We observed significant variations in the manipulation techniques deployed across countries and platforms: the few controversial content in FB FR was mostly associated to (1) and (2) while techniques (1), (2), (3), and (5) were mostly leveraged in the UK and all strategies were used in the US – from (1) to (5) on Facebook and principally (6) on Instagram. Additionally, fake reshares and screenshots of fabricated public figures’ posts were found to be particular to IG US (also found in FB US but not elsewhere), while false individuals’ testimonies (e.g., ‘my friend working at the CDC said…’) is particular to FB US. From a general perspective, C2 content was found to be significantly more subtle on IG US than anywhere else: claims are more suggestive, often based on multimodal combinations (picture and caption), and satire and parody is less obvious – such as memes presenting rioters with streamers, whose text edited in a non-obvious way. Another typical example is the trend to create parodies of Donald Trump’s Twitter-style posts, which may result in unintentional diffusion of misinformation. In addition to the excuse of casualness, the larger use of video formats makes it harder to detect false claims made within long videos that also include personal opinions and testimonies. \medskip

\vspace{-0.3cm}	

The strategies to spread misinformation by coordinated groups of unauthentic actors are still expanding in technicality to avoid bulk detection \cite{goldstein2021disinformation}. At the content level, however, misinformation innovation does not proceed from technology improvements such as deepfakes – as Brennen et al.\,\cite{brennen2020types} found, we observed that altered reported posts come from low-tech photo and video edits – but language refinement and social cues. This follows the evolution of marketing techniques on social networks, recently illustrated by tobacco companies’ efforts to leverage Instagram influencers’ coolness to advertise e-cigarettes, vapes, and nicotine pouches \cite{chapman2021} or Mike Bloomberg’s strategy to hire Instagram influencers ‘to make him seem cool’ with memes \cite{Noor2020Mike}.  While people do not necessarily believe the false news they share \cite{pennycook2021psychology}, it was observed that repeating a claim increased the perceived truthfulness of it \cite{fazio2019repetition}, which could result in this new kind of ‘grey’ content that is just as harmful as fabricated news. Presenting these posts as frivolous may make them not only harder to detect and moderate but may also increase their virality potential and thus their impact due to their repetitiveness.

\vspace{-0.1cm}	

\section*{A plurality of reporting behaviours allows different leverages for noise reduction}

\subsection*{Identification of four reporting behaviours}

It is certainly conjectural to infer users’ intentions based on their reporting activity. The diversity of reported content, however, supports the hypothesis of a plurality of reporting behaviours associated to distinct goals, which several signals allow us to identify.

\begin{enumerate}
    \item \textbf{Reporting false news to flag it to moderators.} This is the expected use of the reporting feature, and several signals support this hypothesis. For example, 41.8\% of all labelled content pieces in S1 were classified as C, of which 17.3\% were confirmed to be false news by fact-checkers (VM), suggesting an accurate use of the reporting tool by a significant portion of reporters. Comment sections are also being used by a number of users to post comments expressing disbelief (e.g., ‘fake’, ‘fake news’, ‘this is a hoax’), communicate that they have reported a post, and encourage others to do so (e.g. ‘reported!’, ‘this is fake, report it’), often providing links to material that debunks claims from fact-checking websites. Some users even act as ‘super-reporters’, flagging an impressive number of relevant content items. For example, 43 Instagram users from S0 were responsible for more than 1,000 reports each over 90 days (May-July 2020). One user logged 2,962 reports for 60 distinct content pieces, of which 11 were labelled in S1, containing 8 C2, 2 C0, and 1 MS. Another one logged 1,175 reports for 37 distinct items; 10 were in S1, in which 5 were C2, 4 C1, 1 C0, and 8 out of the 10 were confirmed to be misinformation by fact-checkers. \vspace{-0.1cm}	

\item \textbf{Reporting due to disagreement or jealousy to annoy the content creator. }The significant number of O content, especially in FB US (22.3\%), suggests that many users report opinions they disagree with and news they believe but dislike. This is consistent with previous research on Instagram \cite{Grossman2020,Smyrnaios2020}. In addition, a significant portion of reported items do not even contain a claim, nor do they qualify for other policy violations. This suggests that reporting is not only used as a ‘dislike button’, namely to send negative feedback to users expressing divergent opinions, but also to annoy them, perhaps to express jealousy resulting from negative social comparison. This hypothesis is suggested by the great number of I content, of which many items are related to romantic relationships (pictures of couples, often with captions expressing love and happiness, or public notifications such as ‘X is in relationship with Y’) and body image (selfies at the gym), two top topics known for triggering negative social comparisons \cite{burke2020social}. As a limitation to this a priori irrelevant reporting, it should be noted that a number of I items were associated to other kinds of user-level offenses (e.g., fake account, impersonation, property rights). This applies to 7.5\% of I content in S1, but it may even apply to a larger number of reported items for which the violation has not yet been detected.
\vspace{-0.1cm}	
\item \textbf{Reporting inappropriate content that is not misinformation misinformation.} All the M content – 23\%, of which 55\% was confirmed policy-breaking by moderators (what we label VO) supports the hypothesis that users ‘misreport’ a number of items. While this content may be problematic, users select the wrong option. M1 content, especially on Instagram, often contains photos that are sexually suggestive or shows quasi-nudity, without however qualifying as pornography or sexual solicitation. M2 content often includes expressions of brutality and MS content refers to scams, spams, and fake accounts. It is understandable that scams and unauthentic accounts could be associated to false information although dedicated categories exist to specifically report these. This supposed ‘mistake’ is however more surprising for M1 and M2 content because to report a post as false news on Instagram a user has to select ‘inappropriate’ instead of ‘spam’, then scroll down to the ‘false news’ option, which is the last one in a list of eight categories which better fit all types of items classified as M1/M2.
\vspace{-0.1cm}	
\item \textbf{Reporting to draw the moderators’ attention to a problematic situation.} M3 items are posts soliciting the user community to support or be aware of an issue which is not primarily political and usually has a personal connexion to the content sharer. It varies from warnings (reporting scams and bad experiences with businesses or artisans), to unidentified accusations (e.g., ‘the waiter of company X was racist to me’), to identified exposure (‘this man is a racist, expose him’ accompanied with screenshots of private messages), often calling for public shaming and sometimes including serious indictments (‘X sexually assaulted me last week’). We find these accusations in every country (51 in FB US, 74 in FB UK, 60 in FB FR), but it is hard to tell whether reporters aim to have moderators take action against the content creator (a few of these turned out to be associated with harassment or created by false accounts), to flag a danger associated to the public shaming of a potentially innocent person, or to support the post sharer in hopes that moderators might alert the police. This latter hypothesis could help understand why when users explicitly ask the community for support (‘please block X’, ‘please report X’, ‘report this false news’), users often report the whistle-blower instead of the post or user they are asked to flag [50,51,52].
\end{enumerate}

The absence of other typologies of reporting behaviours does not allow us to compare this typology with other ones. Therefore, we suggest that these categories are considered a first draft of a typology of reporting behaviours, which would benefit from further research using different methods, notably psychometrical analysis and sociological interviews.

\begin{figure*}[b]
\centering
\includegraphics[width=1\linewidth]{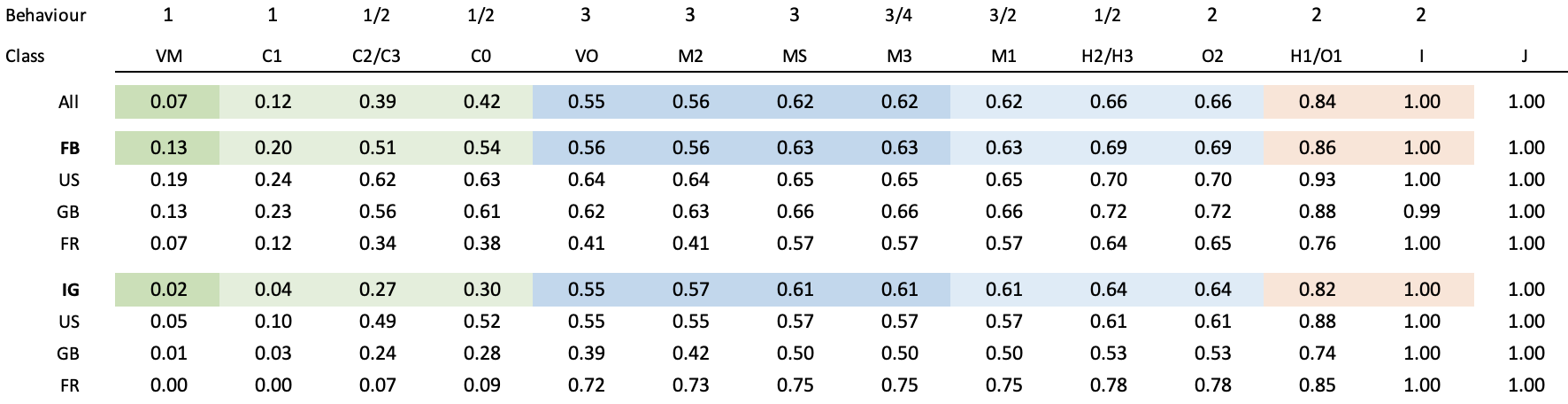}
\caption{Cumulative proprortion of content items per class associated to reporting behaviours.
\textit{This table should be read from left to right as such: before C1 is the sum of all VM content and the remaining C1 content.}}
\label{fig:fig4}
\end{figure*}

\subsection*{Splitting the noise to decrease reporting inaccuracy} 

The GCRC has been shown relevant to analyse users' reports and motivations; we shall now explain that it can also help content moderators better analyse such signal for detection purpose. To this end, we re-order the GCRC classes according to the moderators' interest as in Fig. 4. Moderators care about the relevance of reports from a misinformation moderation perspective, focusing on the signal's accuracy. The traditional approach, calculating the ratio VM/N(S1) gives us a 0.93 \emph{inaccuracy} of the reported content, which explains why user reports are often considered very 'noisy' and reporters untrustworthy. This metric is however subject to several limitations including the fact that the number of S1 items reviewed by fact-checkers is unknown. The identification of reporters' behaviours then allows us to re-evaluate the accuracy of user reporting and assess its performance using more suitable metrics. Such an approach consists in splitting this ‘relative’ inaccuracy into different types of noise, which can be linked to various reporting behaviours and upon which distinct actions can be taken to improve the overall signal. \vspace{-0.1cm}

\begin{itemize}
    \item $False \, noise = \frac{\sum(C1, C2, C3, C0, C2^*)}{N(S1)}$ indicates the proportion of coherent, although not necessarily accurate, reporting. It accounts for 0.35 of the overall inaccuracy and could be associated to the reporting behaviour 1. We call it false noise as it probably mainly results from credible reporting done by people with a low capacity to detect false news or understand the moderation policies and types of misinformation prioritised by fact-checkers (e.g., C0). The use of educative campaigns, communication efforts explaining platforms’ moderation policies, and self-fact-checking material could help reduce this noise.  \vspace{-0.1cm}
    
	\item $Quasi-noise = \frac{\sum(VO, M2, MS,M3)}{N(S1)}$	indicates the proportion of relevant, although not necessarily coherent, reporting. It accounts for 0.2 of the inaccuracy and could be associated to the reporting behaviour 3. We call it quasi-noise because it refers to content which can be moderated but not in the context of misinformation. The reporting is most probably credible, and additional investigations should be conducted to understand the source of the confusion surrounding the reporting feature. \vspace{-0.1cm}

	\item $Soft \, noise = \frac{\sum(M1, H2,H3,O2)}{N(S1)}$
	indicates the proportion of doubtful, although not necessarily irrelevant, reporting. It accounts for 0.03 of the inaccuracy and could be associated to any reporting behaviour. We call it soft noise because it is difficult to make any strong assumptions about it. \vspace{-0.1cm}

	\item $Hard \, noise = \frac{\sum(H1, O1, I, J)}{N(S1)}$
	indicates the proportion of probably irrelevant reporting. Accounting for 0.34 of the inaccuracy, it could be associated to the reporting behaviour 2. We call it hard noise because it most probably results from unfaithful reporting which should be filtered.

\end{itemize}

As we interpret it, around 55\% of the reporting inaccuracy (false and quasi-noise) could probably not be attributed to the reporters’ lack of seriousness but rather to a confusion surrounding the reporting features and moderating rules. While our RBI assumption inflates the number of C2 content, categorising content as ‘false noise’ when it actually is unfaithfully reported, a number of C items have, however, not been reviewed by fact-checkers and could then have been rated VM. Likewise, a portion of the ‘hard noise’ may also contain undetected policy-breaking items, which may be rated VO later on. We are now proposing a way to classify these different types of reporting noise which could potentially be used to redirect feedback towards the relevant moderation channels.

\subsection*{Leveraging multi-channel reporting to classify reported content per type}

\begin{figure}[H]
\centering
\includegraphics[width=1.0\linewidth]{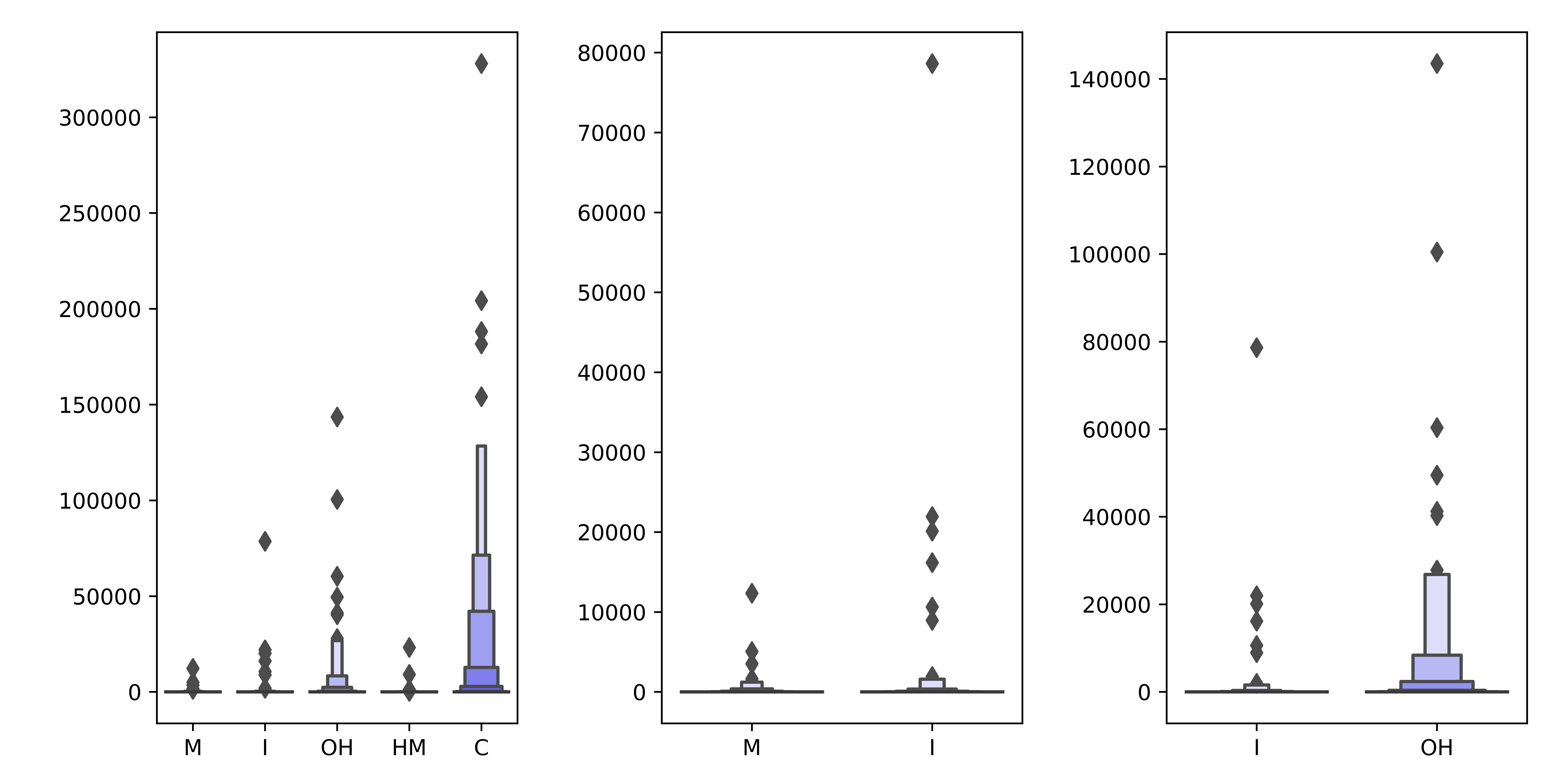}
\caption{Distribution of IG content reports count per aggregated class}
\label{fig:fig5}
\end{figure}

Our objective is to identify the false noise (C), most relevant for misinformation moderation purposes, and the quasi-noise (M), relevant for another moderation channel. We focus on Instagram as it has been shown to be the platform in which new techniques are appearing. To better match the behaviours and noise types previously identified, we aggregate classes as such: C (C0, C1, C2, C3, C2*), M (M2, M3, MS), HM (H2, H3, M1), OH (H1, O1, O2), and I. Figure \ref{fig:fig5} represents the distribution per aggregated class of the number of reports received for each Instagram item of S1 over 90 days. We considered a 0.999 quantile, excluding 4 outliers (1 C, 1 OH, 1 M, 1 I) with a number of reports superior to 329,279. The total number of reports received for an item clearly appears to be a meaningful signal to distinguish C from other classes, especially from I and M. It, however, seems less accurate to differentiate C and OH under a 20,000 reports threshold, and uncapable of separating M from I.

\vspace{-0.1cm}

\vspace{-0.2cm}

\begin{figure}[h]
\centering
\includegraphics[width=0.85\linewidth]{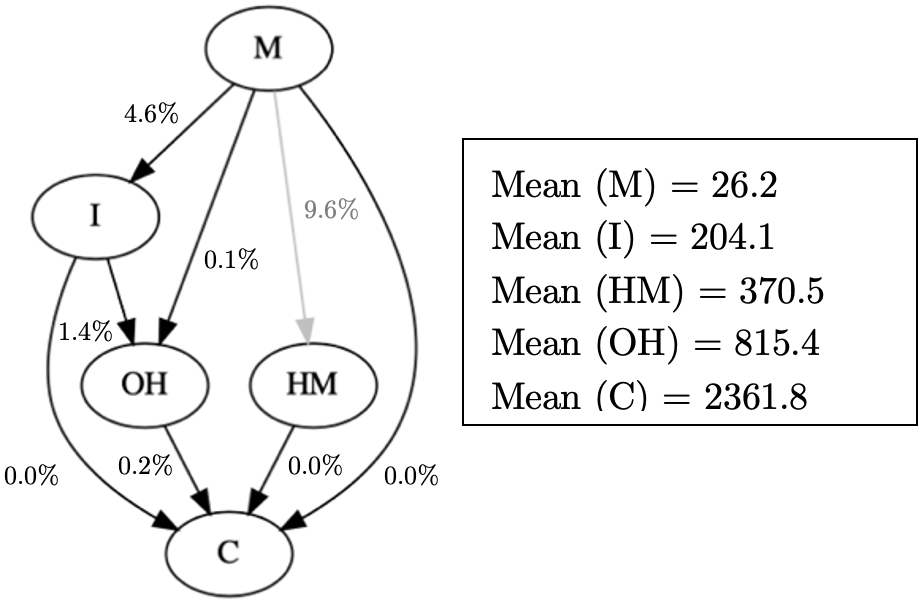}
\caption{Partial order of aggregated classes based on number of reports}
\label{fig:fig6}
\end{figure}

We confirm these observations by comparing the means of the dependent variable for each pair of aggregated class. The results are presented in Figure \ref{fig:fig6}, where the p-values displayed correspond to each two-by-two comparisons. We use Welch's t-test to accept a one-sided alternative hypothesis with significance level of 5\% (or 10\% for grey arrows). We did not assume same variance among classes since the reporting behaviour may significantly vary, and it is empirically verified. The mean normality assumption was verified by the D'Agostino-Pearson test. Kolmogorov-Smirnov tests were also conducted to point significant distribution shift between aforementioned categories.
\vspace{-0.1cm}
\vspace{-0.2cm}

\begin{figure}[h]
\centering
\includegraphics[width=0.9\linewidth]{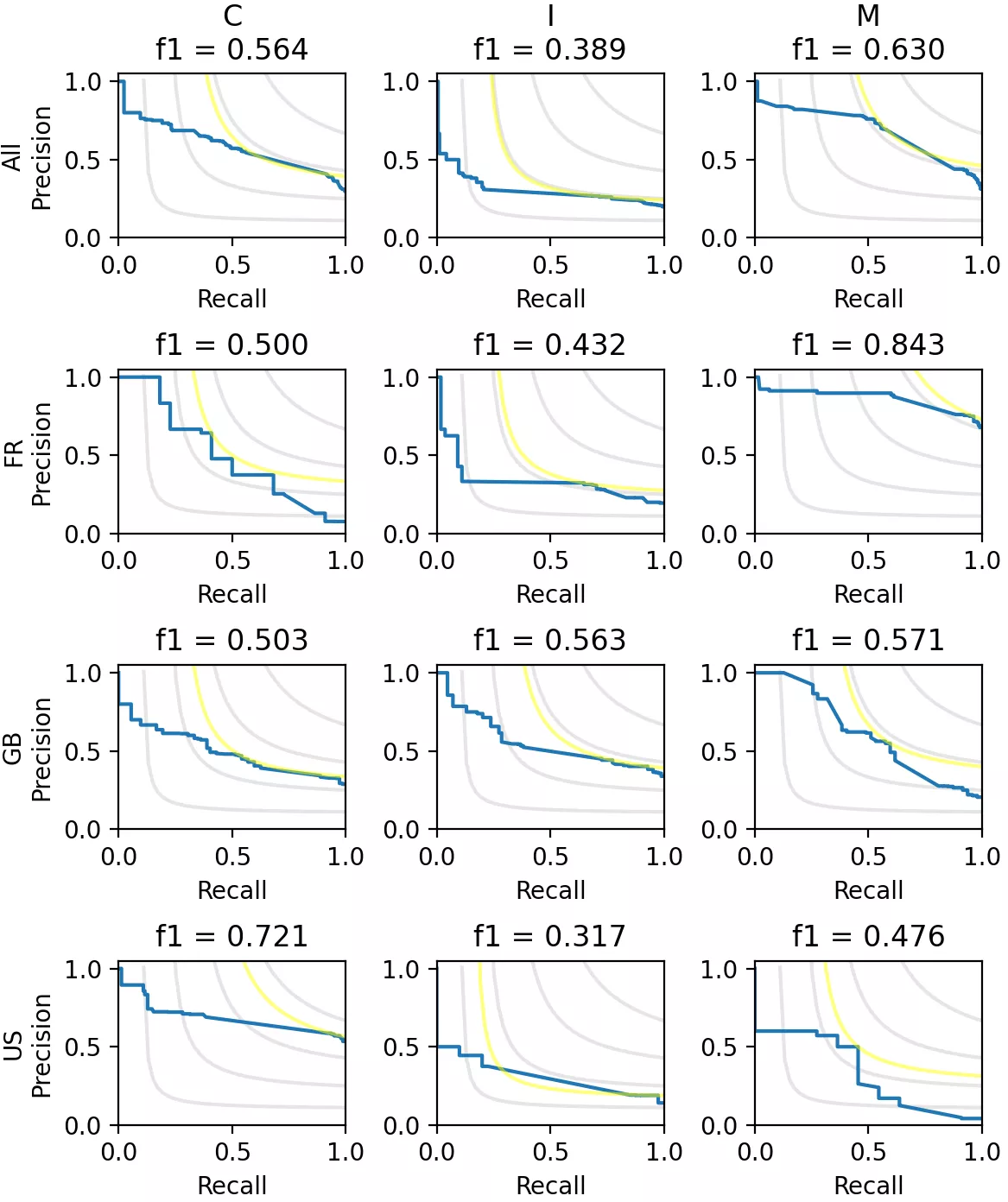}
\caption{Precision-recall curves per country for aggregated classes detection on S1(IG)}
\label{fig:fig7}
\end{figure}

Finally, we train a gradient boosting classification model to identify four classes (C, M, I, others) from ten features corresponding to the main platform reporting categories ('false news', 'nudity/sexual solicitation', 'violence', 'harassment', 'suicide/injury', 'spam', 'hate speech', 'unauthorised sales', 'inappropriate content', ‘I don’t like it’) on IG(S1) with test sample of 10\%. The model’s general performance presented in Figure \ref{fig:fig7} reaches an F1 of 0.56 on C/¬C and of 0.63 on M/¬M. More interesting is the performance per country: the model claims an F1= 0.84 on M/¬M for IG FR, where spam was identified as the main issue, and an F1 = 0.72 on IG US, where misinformation was identified to be the main issue. While they significantly vary in order between countries, the most important features are 'false news', 'spams', 'hate speech' and 'inappropriate content'. This simple model shows that it is possible to learn to classify reported content according to our typology.

\vspace{-0.2cm}

\section*{Conclusion}

By studying content reported as misinformation, comparatively between regions and platforms, and at the content-level, our goal was to refute the idea that user reports are a low-accuracy signal that would not be not very suitable for online misinformation detection. Instead, we show that user reporting offers a complex signal, composed of different feedback that should be understood and assessed separately. The content review allowed for us to observe meaningful distinctions regarding content reporting between countries and platforms as there was significant variation in the volume, type, topic, and manipulation technique. Two key findings are the quasi-absence of C content on IG FR, which instead has a spamming issue, and the apparent convergence between Instagram and Facebook in the US. This uniformization in volumes and topics is accompanied by the emergence of a new type of manipulative technique which is both harder to detect and moderate, raising important challenges for misinformation management. The fact that misinformation innovation is not driven by sophisticated techniques such as deepfakes but rather by language refinement and social cues supports the idea that leveraging users’ support for a more participative online content moderation is an interesting direction to improve misinformation detection in the future. We show here that examining the variety of behaviours present in a given signal can help identify other relevant data points that can be combined to increase the signal’s overall quality. We also show that the typology we propose is predictable, and that a basic model trained a small dataset (n = 4,056) and relying only on a few basic data points can reach a significant performance at detecting its main classes.This may now open perspective to improve the performance of detection algorithms, not only by filtering the reporting signal (based on the different behaviours identified and users' credibility), but also as a feedback within a communication framework (by investigating further the reasons why well-meaning users seem to misreport a number of items).

\subsection*{Author Affiliations}

Facebook AI Research, 9 rue Ménars, 75002 Paris
Ecole Normale Supérieure, Department of Philosophy, 45 rue d'Ulm, 75005 Paris

\acknow{We would like to thank Joelle Pineau, Antoine Bordes and Jerome Pesenti for their support in having this paper published.}

\showacknow{} 

\bibliography{pnas-sample}

\begin{thebibliography}{10}

\bibitem{humprecht2020resilience}
Humprecht E, Esser F, Van~Aelst P (2020) Resilience to online disinformation: A
  framework for cross-national comparative research.
\newblock {\em The International Journal of Press/Politics} 25(3):493--516.

\bibitem{cinelli2021echo}
Cinelli M, Morales GDF, Galeazzi A, Quattrociocchi W, Starnini M (2021) The
  echo chamber effect on social media.
\newblock {\em Proceedings of the National Academy of Sciences} 118(9).

\bibitem{wardle20166}
Wardle C (2016) 6 types of misinformation circulated this election season.
\newblock {\em Columbia Journalism Review} 18.

\bibitem{molina2021fake}
Molina MD, Sundar SS, Le T, Lee D (2021) “fake news” is not simply false
  information: a concept explication and taxonomy of online content.
\newblock {\em American behavioral scientist} 65(2):180--212.

\bibitem{innes2020techniques}
Innes M (2020) Techniques of disinformation: Constructing and communicating
  “soft facts” after terrorism.
\newblock {\em The British journal of sociology} 71(2):284--299.

\bibitem{newman2020reuters}
Newman N, Fletcher R, Schulz A, Andi S, Nielsen RK (2020) Reuters institute
  digital news report 2021.
\newblock {\em Reuters Institute for the Study of Journalism}.

\bibitem{altay2019so}
Altay S, Hacquin AS, Mercier H (2019) Why do so few people share fake news? it
  hurts their reputation.
\newblock {\em new media \& society} p. 1461444820969893.

\bibitem{barthel2016many}
Barthel M, Mitchell A, Holcomb J (2016) Many americans believe fake news is
  sowing confusion.
\newblock {\em Pew Research Center} 15:12.

\bibitem{humprecht2019fake}
Humprecht E (2019) Where ‘fake news’ flourishes: a comparison across four
  western democracies.
\newblock {\em Information, Communication \& Society} 22(13):1973--1988.

\bibitem{Moullot2020}
Moullot P, Halissat I (2020) Masques : comment le gouvernement a menti pour
  dissimuler le fiasco.
\newblock {\em Libération}.

\bibitem{Yurkova2018}
Yurkova O (2018) Six fake news techniques and simple tools to vet them.
\newblock {\em Global Investigation Journalism Network}.

\bibitem{goldstein2021disinformation}
Goldstein~Josh A, Grossman S (2021) How disinformation evolved in 2020,
  brookings tech stream, january 4.

\bibitem{brennen2020types}
Brennen JS, Simon FM, Howard PN, Nielsen RK (2020) Types, sources, and claims
  of covid-19 misinformation.

\bibitem{chapman2021}
Chapman M (2021) New products, old tricks? concerns big tobacco is targeting
  youngsters.
\newblock {\em The Bureau of Investigative Journalism}.

\bibitem{Noor2020Mike}
Poppy N (2020) Mike bloomberg will pay you \$150 to say nice things about him.
\newblock {\em The Guardian}.

\bibitem{pennycook2021psychology}
Pennycook G, Rand DG (2021) The psychology of fake news.
\newblock {\em Trends in cognitive sciences}.

\bibitem{fazio2019repetition}
Fazio LK, Rand DG, Pennycook G (2019) Repetition increases perceived truth
  equally for plausible and implausible statements.
\newblock {\em Psychonomic bulletin \& review} 26(5):1705--1710.

\bibitem{Grossman2020}
Grossman S, et~al. (2020) Reporting for duty: How a network of pakistan-based
  accounts leveraged mass reporting to silence critics.
\newblock {\em Stanford Internet Observatory Cyber Policy Center}.

\bibitem{Smyrnaios2020}
Smyrnaios N, Papaevangelou C (2020) Le signalement sur les réseaux sociaux, un
  moyen de modération mais aussi de censure.
\newblock {\em La revue des médias}.

\bibitem{burke2020social}
Burke M, Cheng J, de~Gant B (2020) Social comparison and facebook: Feedback,
  positivity, and opportunities for comparison in {\em Proceedings of the 2020
  CHI Conference on Human Factors in Computing Systems}.
\newblock pp. 1--13.

\bibitem{Rashkin2017truth}
Rashkin H, Choi E, Jang JY, Volkova S, Choi Y (2017) Truth of varying shades:
  Analyzing language in fake news and political fact-checking in {\em
  Proceedings of the 2017 conference on empirical methods in natural language
  processing}.
\newblock pp. 2931--2937.

\end{thebibliography}

\newpage 

\section*{Supplementary materials}

\subsection*{Annex 1.A: Composition of the data sample S1}

From all the content reported as ‘false information’ by Facebook and Instagram users from France, the UK, and the US between June 3rd and July 3rd, 2020, we extracted a subset S0 that contained all items that included data relevant to our study – i.e., basic information related to content, reporters, and sharers. Because Instagram has only recently allowed people to register their gender, S0 only contains items reported by Instagram users who have linked their account to a Facebook account with a registered gender. From S0, we extracted the sample S1, which is composed of randomly selected items that ensure a relative balance between reporters’ countries, genders, and age categories while maximising the diversity of content and reporters (i.e., minimising the number of identical posts shared or reported by different users and that of different items reported by the same user). S1 is composed of 11,463 content pieces, of which 8,975 received a class label, 2,004 were removed by users before being labelled, and 484 were not published in French or English. Among all labelled content, we shall refer to S1 subsets as follows: FB FR (n = 1,422), FB UK (n = 1,519), FB US (n = 1,527), IG FR (n = 1,491), IG UK (n = 1,516), and IG US (n = 1,500). Although the subsets’ sizes may seem small compared to S0’s (which contains 97.2 million entries for Instagram reports alone), only 0.23\% of IG S0's content items are distinct, compared to 97\% in IG S1. This results in the 1,491 labelled items of IG FR representing 6.3\% of all the different items reported on Instagram France in S0.

\subsection*{Annex 1.B: Review methodology}

The review process was composed of four steps. As there was no existing typology for reported content, an exploratory review of c. 300 items per country allowed us to lay the foundations for the classification. Such a review was conducted independently by the authors, who then compared their results to define the final version of the GCRC. We then reviewed each of S1’s items, associating them with a GCRC class. The process was ordered to review content from the same media type, then platform, then country together to assess differences and similarities across these three layers. Every time that all items from a given platform in a given country were labelled, a confirmation review was conducted per content class to ensure the coherence of the labels. Once all items were classified, a final review was conducted per class labels across all content. In addition to detecting misclassified items and ensuring the coherence of the labelling across platforms and countries, this final round also permitted us to include fact-checkers’ most recent ratings (as of December 7th, 2020). The resulting classification, presented below, contains six classes and fourteen sub-classes. Note that it does not aim to assess the reporters’ performance – that is, distinguishing reported content containing accurate vs. false claims – but credibility – differentiating content that could contain false news from that which could not. This, in addition to the reporter-oriented approach irigger builds on, makes it relevant from a research perspective but should not be used as a turnkey tool to operationalise content moderation.

While we recognise that the fact for the review to have been conducted by the authors may be seen as a methodological limitation, we think that this is justified by the particular difficulty of the labelling task and the granularity of the classification. To be clear, we consider that such a task requires researchers familiar with misinformation topics and literatur and that it could not have been performed in good conditions by a variety of non-specialised annotators. This is also the reason why we elaborate extensively below on the challenges of such a task and the mythological choices they resulted in. 

\subsection*{Annex 1.C: Four difficulties in labelling content reported as misinformation}

Labelling content from a misinformation viewpoint is a complex task which cannot be considered an exact science. It requires someone to (1) identify a claim within a structured piece of content. If it is explicit, they must (2) determine whether it could theoretically be falsified and how difficult the verification process would be in practice. When it is implicit, they must (3) assess the underlying claim’s degree of obviousness. Finally, (4) the difference between the subjective perception of the content sharer and the reporter should also be considered. We identified four main difficulties associated to these steps. \medskip

The first difficulty is design-specific, resulting from the content’s structure. Some items are multi-layer posts, composed of content pieces (e.g., a link, picture, or video), meta-content (e.g., a caption or edits of the video) and second-level meta-content (e.g., a caption reacting to a reshared post, which already included a caption), making the identification of the reported claim difficult. To address this, we adopted a holistic reporter-oriented approach, holding a reporter’s best intention (RBI) assumption: if any of the possible claims expressed in any layer of the item could be reasonably considered controversial, we assumed the user reported the item because of this claim. We herein aimed to minimise the number of false negatives – items labelled as irrelevant whereas reporters did try to report accurately – at the cost of false positives. Consistent with our position of assessing the content’s falsifiability and not its veracity, there is a high ratio of C2 classes among shared links on Facebook (c. 40\%) as articles’ headlines often contain factual claims. The RBI, however, allowed us to realise that many posts that seemed irrelevant at first did not derive from frivolous reporting but contained detected policy violations other than misinformation. \medskip

The second difficulty relates to a claim’s degree of falsification, as two claims may be equally true or false but differ in the resources needed to verify them. Consistent with our reporter-oriented approach, we adopted a best fact-checking resources (BFR) assumption, considering all identified controversial claims as relevant whatever the resources required to verify them. Reporters have little information about the fact-checkers’ capacities, which should then be orthogonal to their reporting credibility. While we did not aim to rank controversial claims by importance, it was nevertheless possible to draw a relevant distinction between small-scale and large-scale impact news. The impact scale differs from the geographic scale as events with a small geographic scale may have a large-scale impact. Many M2/C2 content reported in the US herein does not necessarily relate to large protests but to local events from which larger associations could be made – e.g., an individual’s act of violence presented as epitomizing the whole Black Lives Matter (BLM) movement. Whereas an event’s geographic scale was often found to have little relevance for assessing its potential impact, a small-scale impact was usually associated to a small geographic scale. As the reporting of such content also seems to proceed from a different intention, we distinguished small-scale impact (C0) from large-scale impact (C1, C2, C3) content.\medskip

The third difficulty is semantic, grounded in the thin distinction between assertion and suggestion. Many items do not explicitly endorse a controversial claim but instead suggest it in various ways. It is even more complicated to assess this content when it includes emojis, multi-modal associations (e.g., on its own, the caption or image is not controversial, but their combination is [49]), or refers to a commonly known idea without a direct reference. The several rounds of reviews allowed us to develop a general understanding of the top viral topics and reclassify the items whose suggestive references had previously been missed. The RBI also allowed us to classify items with a suggested claim as belonging to C1 or C2 according to the obviousness of the reference and the level of controversy.\medskip

The fourth difficulty is metacognitive, resulting from a double asymmetry. The first one is that between the actual intention IA(A) of a user A when posting a post PA and the given intention IB(A) inferred by user B of A when seeing PA. This is particularly the case when A makes a metaphorical use of statistics (e.g., ‘99\% of people recover from Covid-19’; suggesting that the large majority of people recover, which is accurate even though the exact statistics may not be) or hyperboles (e.g., ‘everybody recovers from Covid-19’, suggesting that most people do). The second one affects A’s expected reception of PA by B and B’s actual reception. This especially applies to humoristic posts, understood as such by some people but taken seriously by others. This pitfall is central as it puts an agent’s subjectivity in tension with that of others; whereas we aimed to stick to the RBI, a number of reported items of content clearly expressed irony. However, many humoristic posts also contained a sub-claim that was often controversial, making humourful posts a difficult-to-moderate vector for hoax dissemination. To satisfy the diversity of cases, we broke content using humour into three sub-classes according to the content’s obviousness and potential to offend. Many items also contained a mixture of opinions and fact-checkable news or combined inappropriate elements and controversial claims. We classified the former as C instead of O because mixed posts remain relevant from a misinformation reporting perspective, and the latter as C instead of M, aligned with the RBI.\medskip

Finally, the RBI revealed itself as a solid asset that preserved the labelling process against the reviewers’ personal opinions. In such a context of great uncertainty, it saved us from the temptation of classifying posts as I when they contained claims that seemed obviously accurate, debunked later on, or obviously false but that were ultimately confirmed. Based on the content’s nature, GCRC classes are sufficiently objective to be robust to the plurality of opinions that reviewers may have, while subclasses are more penetrated by reviewers’ opinions. This two-level classification thus combines the advantages of a highly consensual labelling process at the class-level and the integration of meaningful additional signals, that are however less univocal, at the subclass-level. Comparing with fact-checkers’ ratings, we find that 97\% of confirmed false news are rated C. The other 3\% relate to O1 content, for which we disagree with these ratings.

\subsection*{Annex 2: Six manipulative strategies to convey misinformation}

The first two strategies target people with a certain degree of scepticism.

\begin{enumerate}
    \item \textbf{The revelation} technique mostly characterises typical conspiracy theories (C1). It challenges people’s egos and encourages them to ‘wake up’ instead of being a ‘sheep’, making direct references to the cabal, Masons, a world elite, and a new order. These posts are usually marked by semantic patterns either related to the general concepts of truth, trust, the elite, and the establishment (e.g., ‘news’, ‘media’, ‘wake up’, ‘masons’, ‘governments’, ‘truth’, ‘sheep’, ‘facts’, ‘distracting’) or to more contextual theories such as QAnon or anti-vax conspiracies (e.g., ‘pizza’, ‘paedophile’, ‘Hollywood’, ‘children’, ‘Clinton’, ‘Trump’, ‘Bill Gates’, ‘chip’) and intend to gain virality by soliciting viewers to reshare (‘spread the word’, ‘share before it gets deleted’, and ‘share to expose them’). Congruent with other research \cite{Rashkin2017truth}, such findings also confirm that semantic cues can constitute a useful signal to detect misinformation, notably by monitoring the frequency of words contained in posts verified as hoaxes by fact-checkers. This signal, however, might mostly be useful to detect the most caricatured conspiracies on Facebook (C1) such as hoaxes that could be debunked with a quick web search. Moreover, the people most susceptible to fall for these messages certainly are those already in contact with conspiracies theories, and they may not be sensitive to debunking material – being a conspiracist is less a question of being given accurate information than a psychological posture. Semantic methods may also be challenged by counter-detection strategies (e.g., ‘c0r0navirus’).
    
    \item \textbf{The critical tipping point} technique consists in leveraging a real fact (e.g., a public scandal or polemical claims from a controversial figure) as an entry point to encourage people to reconsider all their beliefs. The fact is usually presented in a twisted way, often accompanied by exaggerated empathy (‘this is despicable’) or an invitation to generalise (‘if they lied about this… what else did they lie about?’). A variant consists in creating a false mystery around an accurate fact (‘this is happening, why is nobody talking about it?’). This technique may best work to encourage people who are already sceptical or indignant about a recent scandal to start accepting conspiracies.\medskip

The two following techniques target people who are sensitive to pragmatic arguments.

\item \textbf{False facts supported by false evidence} are typically false statements about events that have allegedly happened and false quotes from public figures, often backed by unauthentic documents (e.g., inauthentic ‘leaked’ documents from the FBI, CDC, or BLM management), unreferenced scientific data, or personal testimonies from a mysterious authority (‘a friend working at the NHS’, ‘the head of the resuscitation department of this hospital’) whose source is impossible to verify. It also includes modified pictures and videos, although we did not find any deepfakes.

\item \textbf{The misleading presentation of facts} consists in presenting authentic documents or accurate facts in a misleading way to encourage a targeted erroneous interpretation (e.g., quotes and pictures taken out of their original context, truncated videos, partial references to history). While similar to the previous one, this technique is harder to debunk because of the accuracy of the facts it is based on.\medskip

The fifth technique plays on people’s empathy.

\item\textbf{The confusion of feelings }consists in presenting either false news or authentic facts in a twisted way to provoke an emotional reaction against a given target. This was particularly observed when someone leveraged an isolated action to discredit a whole movement through the mobilisation of symbols (e.g., protester’s violence against a veteran, acts of police brutality against a peaceful protester, profanation of a military cemetery, destruction of a public figure’s statue). \medskip

The sixth technique plays on the register of coolness.

\item \textbf{The excuse of casualness} characterises items with humour (jokes, ironical statements, memes, caricatures) and/or an artistic dimension (cartoons, short format videos, songs), making more or less explicit reference and reacting to a sub-claim using a frivolous tone. This content differs from both ‘parodies’ (Wardle 2016) and ‘satires’ as it is not ‘meant to be perceived as unrealistic’ (Molina et al. 2019, 198) but rather hides an assumed claim behind a veil of frivolity.
\end{enumerate}

\newpage

\onecolumn

\subsection*{Annex 3: Selected examples of reported content from S1}\,

\begin{figure}[H]
\centering
\includegraphics[width=0.85\linewidth]{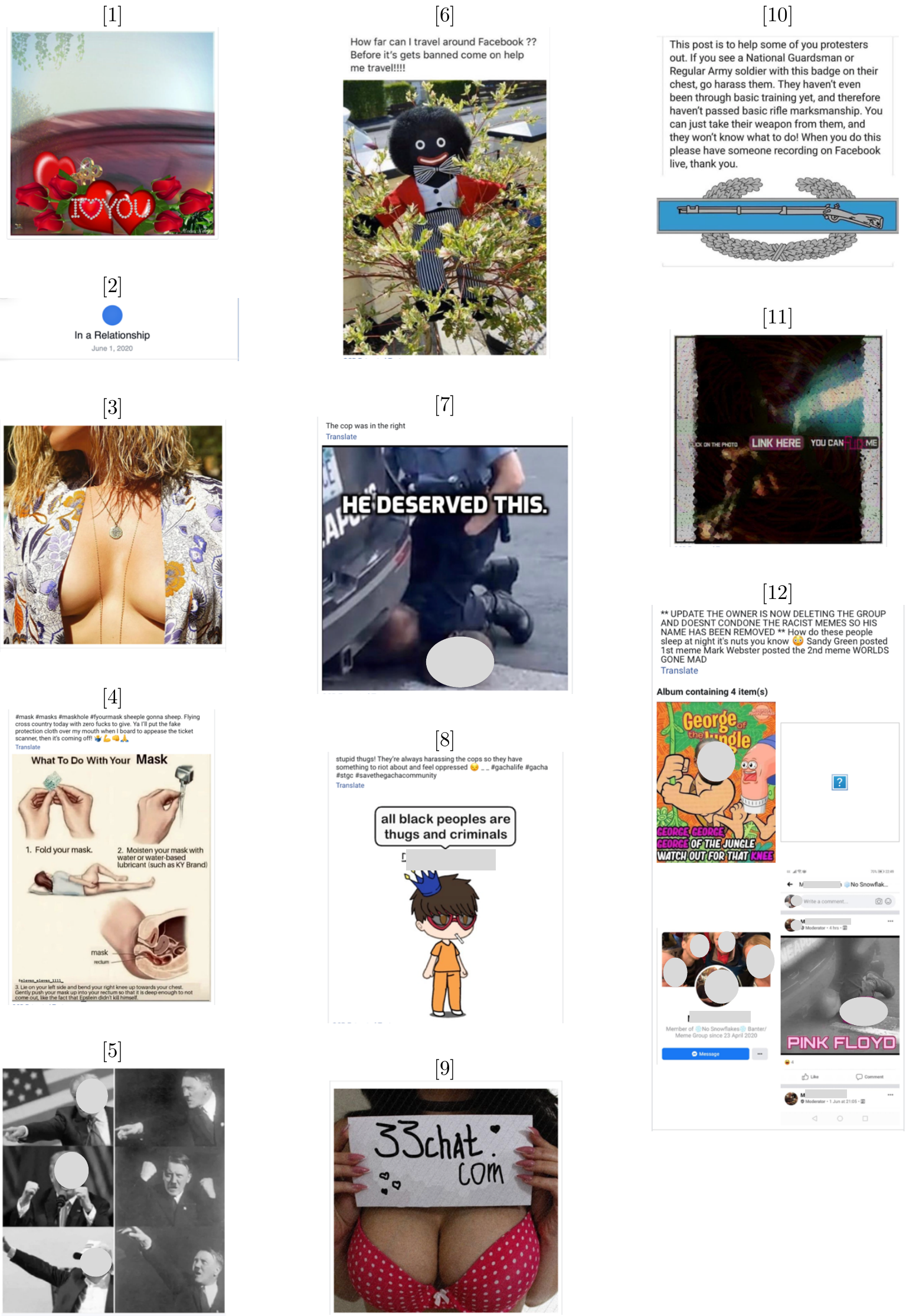}
\end{figure}

\begin{figure*}[h]
\centering
\includegraphics[width=0.85\linewidth]{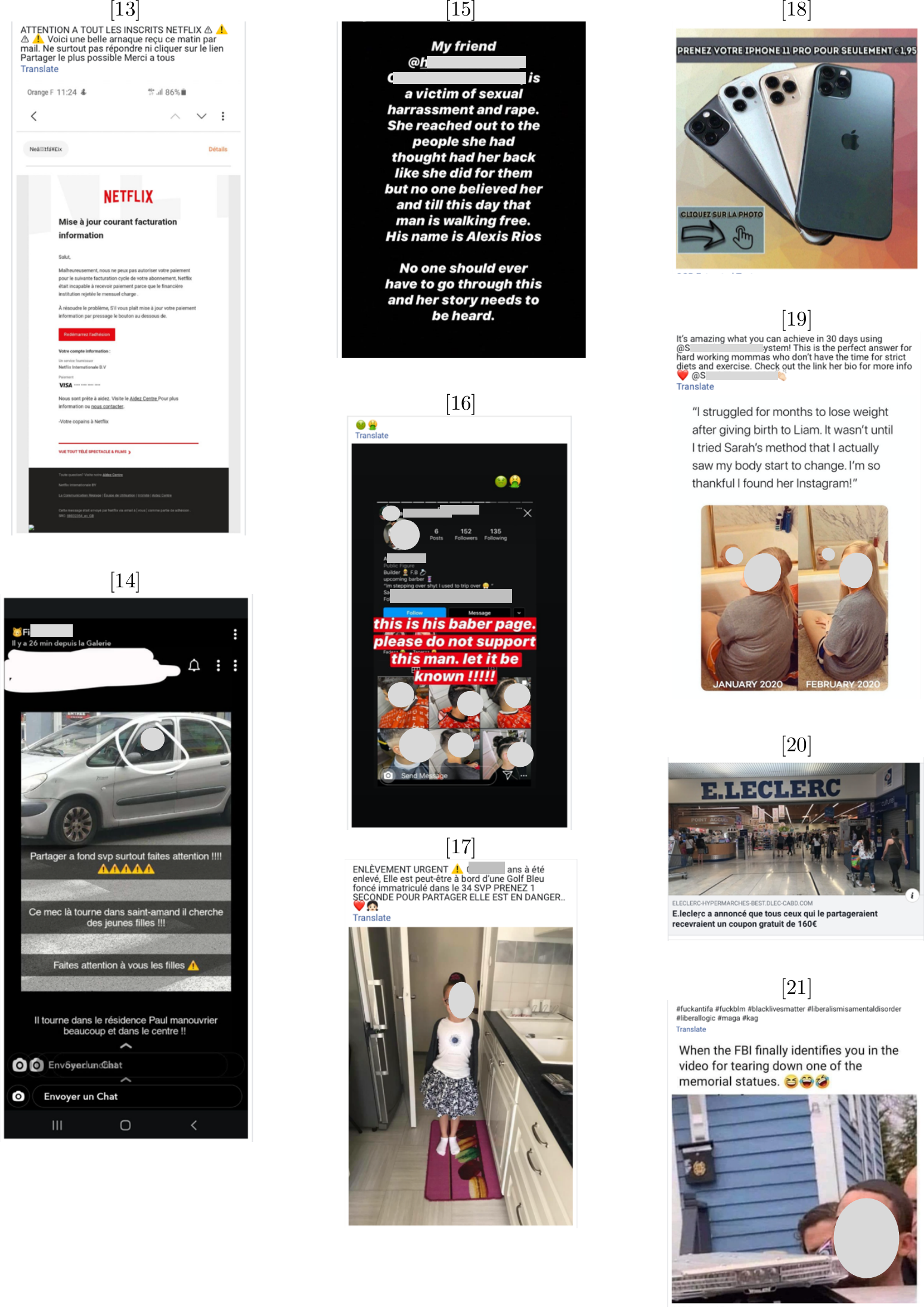}
\end{figure*}

\begin{figure*}[h]
\centering
\includegraphics[width=0.85\linewidth]{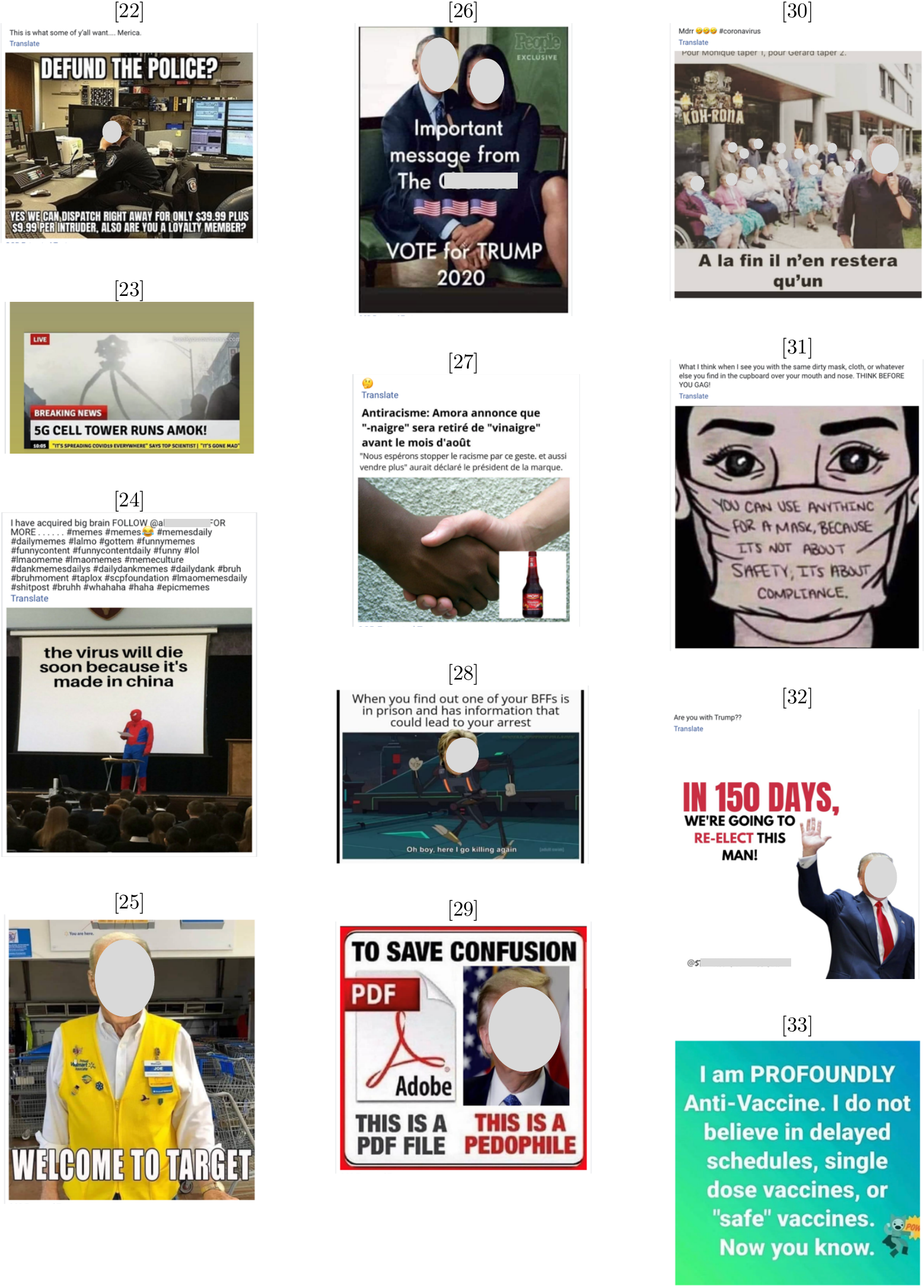}
\end{figure*}

\begin{figure*}[h]
\centering
\includegraphics[width=0.85\linewidth]{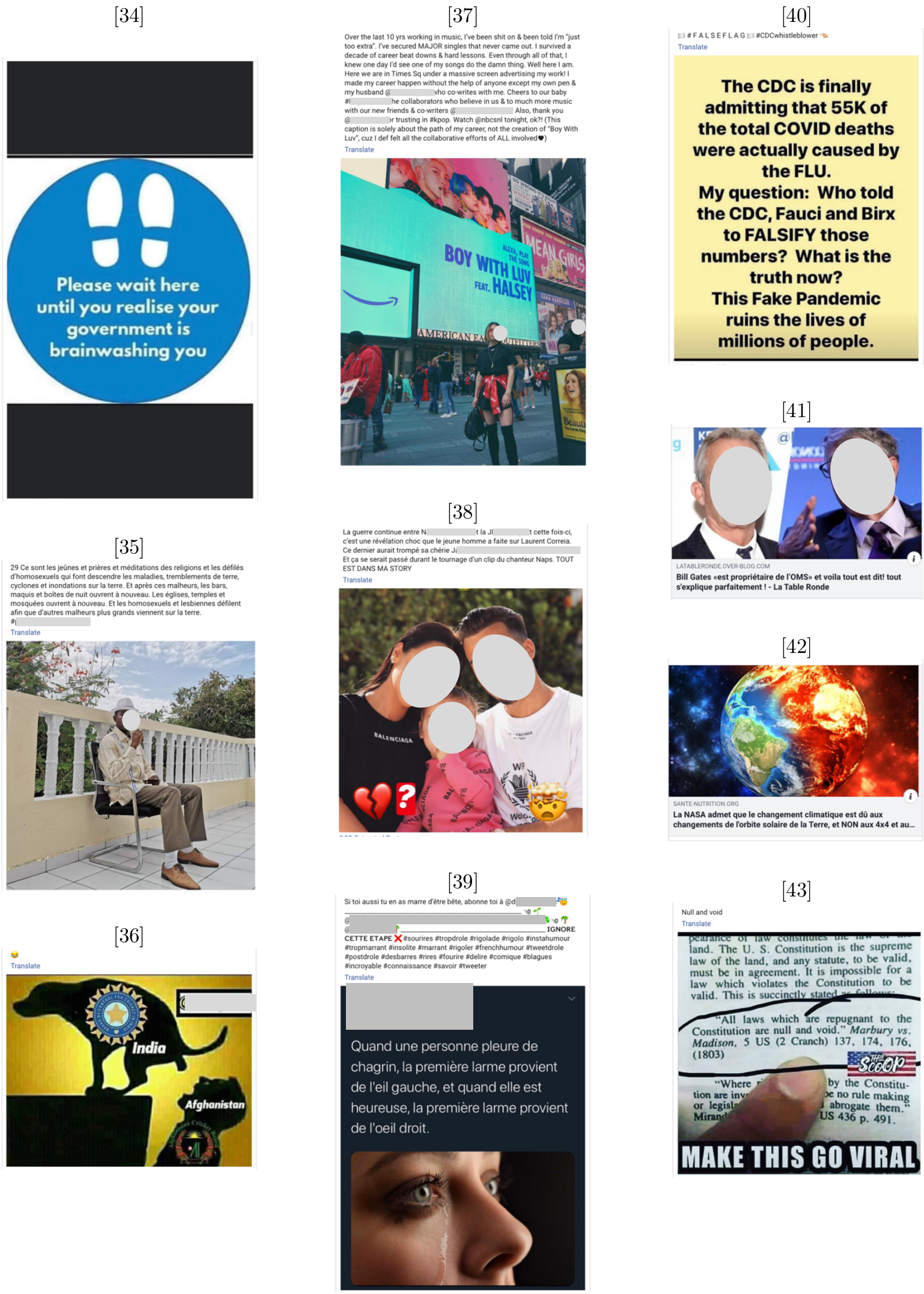}
\end{figure*}

\begin{figure*}[h]
\centering
\includegraphics[width=0.85\linewidth]{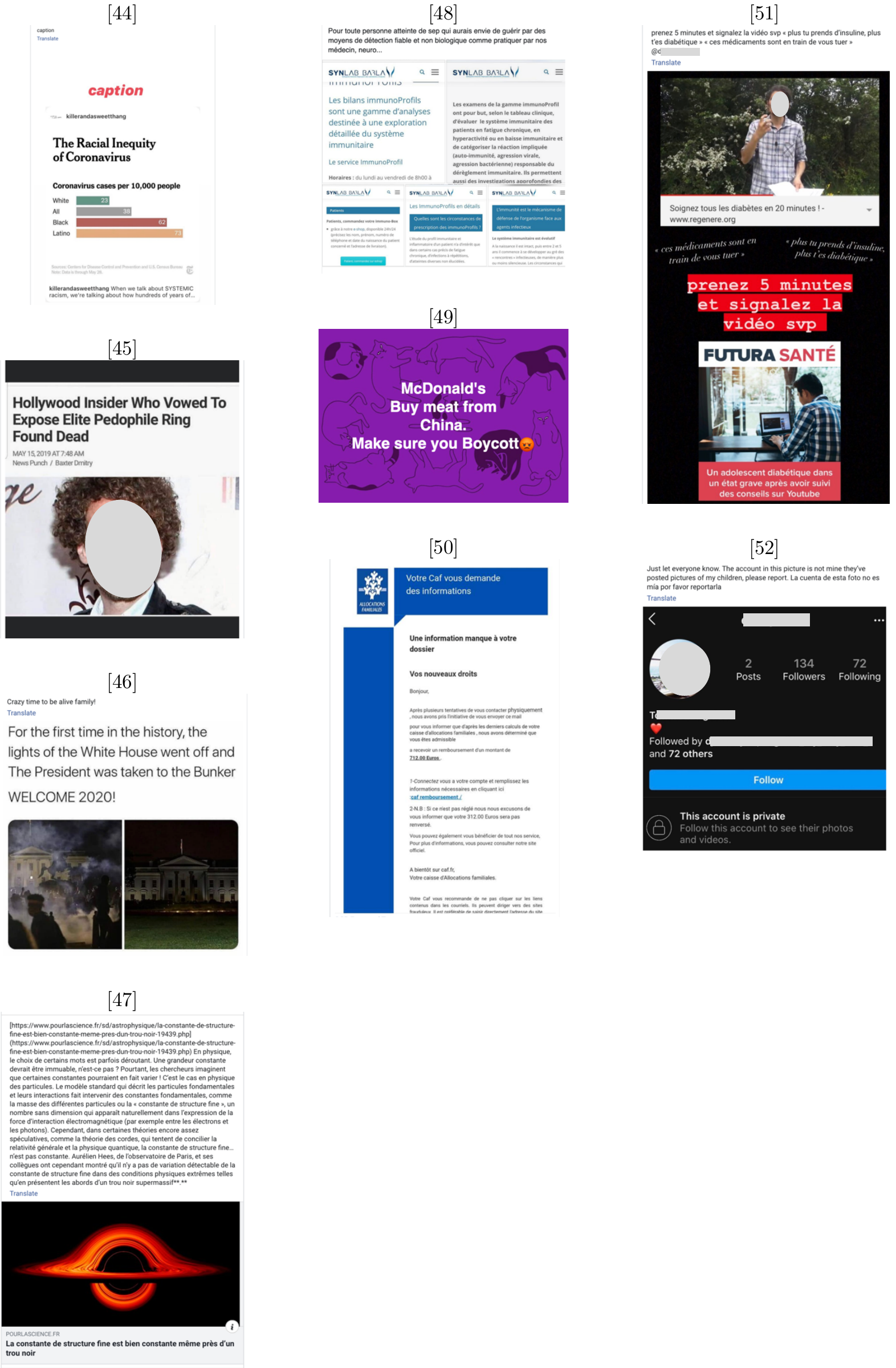}
\end{figure*}

\end{document}